\documentclass{ws-p9-75x6-50}

\newcommand{\bra}{\langle}
\newcommand{\ket}{\rangle}

\begin{document}

\title{The $\mathbf{\Lambda\Lambda-\Xi N}$ Coupling and the Binding
Energy of $\mathbf{_{\Lambda\Lambda}^{\ \,6}}$H\lowercase{e}}

\author{I. R. Afnan}

\address{Department of Physics, The Flinders University of South
Australia\\ G.P.O. Box 2100, Adelaide 5001, Australia}

\maketitle

\abstracts{To understand the difference between the value of the $nn$ and
$\Lambda\Lambda$ two-body matrix elements, we examine the role of the
coupling between the $\Lambda\Lambda$ and $\Xi N$ channels, first 
within the framework of $SU(3)$, and second in the numerical results for
the binding energy of $_{\Lambda\Lambda}^{\ \,6}$He. We find that it is
essential to include the coupled channel nature of the $BB$ interaction
as we proceed from strangeness zero to the strangeness -1 and -2
channels if we are to understand the relative magnitude of the two-body
matrix elements in the different strangeness channels.}

\section{Introduction}\label{sec.1}

The interest in the strangeness $S=-2$ baryon-baryon ($BB$)
interaction was initially motivated by Jaffe's\cite{Ja77} quark model 
prediction of the existence of a flavor singlet $[uuddss]$ state,
known as the $H$ particle. More recently, the interest in the role of 
strangeness in neutron stars, and the possible production of nuclei 
with strangeness $S\le -2$ at RHIC, has given further impetus to
extend the $BB$ interaction to the $S\le -2$ channel.

At present the only experimental data on the $\Lambda\Lambda$
interaction are the three observed $\Lambda\Lambda$ hypernuclei 
$_{\Lambda\Lambda}^{\ \,6}$He,\cite{Pr66} $_{\Lambda\Lambda}^{\
10}$Be,\cite{D63,D89} and $_{\Lambda\Lambda}^{\
13}$B.\cite{A90,D91} These invariably give effective S-wave matrix
elements of
\begin{eqnarray}
\Delta B_{\Lambda\Lambda}(_{\Lambda\Lambda}^{\ \,6}{\rm He}) &=& 
10.9 \pm 0.5 - 6.14 \pm 0.05 \approx 4.7\ {\rm MeV}\nonumber \\
\Delta B_{\Lambda\Lambda}(_{\Lambda\Lambda}^{\ 10}{\rm Be}) &=&
17.7 \pm 0.4 - 13.42 \pm 0.08 \approx 4.3\ {\rm MeV}\label{eq:1} \\
\Delta B_{\Lambda\Lambda}(_{\Lambda\Lambda}^{\ 13}{\rm B}) &=&
27.5 \pm 0.7 - 11.69 \pm 0.12 \approx 4.8\ {\rm MeV} \ ,\nonumber
\end{eqnarray}
which suggests that the matrix element of the $\Lambda\Lambda$
interaction in light nuclei is $- \bra V_{\Lambda\Lambda}\ket 
\approx 4 - 5$~MeV. This value is to be compared with the
corresponding $nn$ matrix element $- \bra V_{nn} \ket \approx
6-7$~MeV, and the $\Lambda N$ matrix element $- \bra V_{\Lambda N}
\ket \approx 2-3$~MeV. As a result we have that $-\bra V_{nn} 
\ket > - \bra V_{\Lambda\Lambda} \ket > - \bra V_{\Lambda N}
\ket$. The question we would like to address in the present report
is: Can we understand this relative magnitude of the $S=0,-1,-2$
matrix elements in light nuclei in terms of flavor $SU(3)$ at the 
meson-baryon level as given in a one-boson-exchange (OBE) potential? 
Or do we need to include explicit quark-gluon degrees of freedom? 
We will demonstrate that the relative magnitude of the matrix 
elements in the $S=0,-1,-2$ can be understood only if we include 
the coupling between the $\Lambda N$ and $\Sigma N$ in the $S=-1$ 
channels, and the $\Lambda\Lambda$ and $\Xi N$ in the $S=-2$ channel.

\section{The Coupled Channel Problem in Flavor $\mathbf{SU(3)}$}\label{sec.2}

Let us assume that flavor $SU(3)$ is a perfect symmetry of strong 
interaction, then the baryon octet ($B=N,\Lambda,\Sigma, \Xi$) 
forms a degenerate multiplet, and the $BB$ system will be in one of the 
six irreducible representations of $\{8\}\otimes \{8\}$, i.e.
\begin{equation}
\{8\}\otimes \{8\} = [\ \{1\}\oplus \{8_s\} \oplus \{27\}\ ] \ 
\oplus \  [\ \{8_a\} \oplus \{10\} \oplus \{\bar{10}\}\ ] \ ,\label{eq:2}
\end{equation}
where $\{8_{a}\}$, $\{10\}$ and $\{\bar{10}\}$ are the flavor
anti-symmetric channels, while the $\{1\}$, $\{8_{s}\}$ and $\{27\}$ are 
the flavor symmetric channels. The flavor wave function in the
$^1$S$_0$ partial wave for the $nn$, $\Lambda N$ and $\Lambda\Lambda$ 
channels can now be written with the help of the $SU(3)$ 
Clebsch-Gordan coefficients,\cite{deS63} as
\begin{eqnarray}
|nn\ket &=& |\{27\}\ket \nonumber \\
|\Lambda N\ket &=& \sqrt{\frac{36}{40}}\ |\{27\}\ket - 
                      \sqrt{\frac{4}{40}}\ |\{8_s\}\ket  \label{eq:3} \\
|\Lambda\Lambda\ket &=& \sqrt{\frac{27}{40}}\ |\{27\}\ket -
                           \sqrt{\frac{8}{40}}\ |\{8_s\}\ket -
                           \sqrt{\frac{5}{40}}\ |\{1\}\ket \ .\nonumber
\end{eqnarray}
Here the $\{27\}$ representation is the only component of the $nn$ 
flavor wave function. Since the $nn$ interaction in the $\{27\}$
representation is strongly attractive to the extent that it almost 
supports a bound state, and both the $\Lambda N$ and $\Lambda\Lambda$ 
wave flavor wave functions have their largest component in the 
$\{27\}$ representation, we expect the $\Lambda N$ and $\Lambda\Lambda$ 
interactions in the $^1$S$_0$ to also be dominated by the $\{27\}$ 
representation. As a result the interactions in the $S=0,-1$ and $-2$ 
are expected to satisfy the relation
\begin{equation}
- \bra V_{nn}\ket > - \bra V_{\Lambda N} \ket > 
                    - \bra V_{\Lambda\Lambda} \ket   \ . \label{eq:4}
\end{equation}
This is in contradiction with the experimental observation from the 
three $\Lambda\Lambda$ hypernuclei which requires that $- \bra 
V_{\Lambda\Lambda} \ket > - \bra V_{\Lambda N}\ket$. The 
inclusion of the $\{8_{s}\}$ and $\{1\}$ component of the flavor wave 
function into our analysis may not improve the situation since there 
are indications from OBE potentials that the interaction in these two 
channels are repulsive\cite{R92} which would reduce the 
$\Lambda\Lambda$ matrix element more than the $\Lambda N$ matrix element.

Since flavor $SU(3)$ is broken in nature to the extent that the masses 
of the $N$, $\Lambda$, $\Sigma$ and $\Xi$ are not equal, the $BB$ 
problem in the $S=-1,-2$ reduces to a coupled channel problem with two 
or more thresholds in each strangeness channel. Thus for $S=-1$ we have the 
$\Lambda N$ and $\Sigma N$ thresholds which are separated by some 
80~MeV; for the $S=-2$ we have the $\Lambda\Lambda$, $\Xi N$ 
and $\Sigma\Sigma$ thresholds with the separation between the 
$\Lambda\Lambda$ and $\Xi N$ channels being $\approx 25$~MeV. This 
suggests that the coupling in the $S=-2$ may be more important 
than is the case in the $S=-1$ channel. To examine the effect of the 
baryon mass splitting on $\Lambda N$ and $\Lambda\Lambda$ matrix 
elements, we introduce effective potentials that include the 
coupled channel effects. These potentials, to lowest order in the 
coupling, are given by
\begin{eqnarray}
\bra V_{\Lambda N}^{\rm eff}\ket &\approx& \bra V_{\Lambda N}\ket - 
            \frac{|\bra \Lambda N|V|\Sigma N\ket|^2}{\Delta E_{YN}}
            \ ;\qquad \ \Delta E_{YN}\approx 80\,\mbox{MeV}\nonumber\\
		   &&                                   \label{eq:5}\\
\bra V_{\Lambda\Lambda}^{\rm eff}\ket &\approx& 
                   \bra V_{\Lambda\Lambda}\ket - 
                   \frac{|\bra\Lambda\Lambda |V|N\Xi\ket|^2}
                        {\Delta E_{\Lambda\Lambda}}\ ; 
    \qquad\Delta E_{\Lambda\Lambda}\approx 25\,\mbox{MeV}\ .\nonumber
\end{eqnarray}
The fact that the thresholds in the $S=-2$ channel are closer than in 
the $S=-1$ channel suggests that we get more attraction in the $S=-2$ 
effective interaction due to the coupled channel nature of the problem, 
than in the $S=-1$ channel. If in addition we assume that the matrix 
elements that determine the strength of the coupling can be extracted 
from $SU(3)$, i.e. flavor $SU(3)$ is approximately a good symmetry, 
then
\begin{eqnarray}
\bra\Lambda\Lambda|V|\Xi N\ket &=& -\frac{18}{40}V_{\{27\}} +
\frac{8}{40}V_{\{8_s\}} + \frac{10}{40}V_{\{1\}} \nonumber \\
   & &                                               \label{eq:6}\\
\bra\Lambda N|V|\Sigma N\ket &=& -\frac{12}{40}V_{\{27\}} + 
\frac{3}{40}V_{\{8_s\}}\ . \nonumber
\end{eqnarray}
If the interaction in the $\{27\}$ representation is dominant, then the
coupling between the $\Lambda\Lambda$ and the $\Xi N$ channels is 
stronger than is the case between the $\Lambda N$ and the $\Sigma N$ 
channels. This again supports the argument that the coupling resulting 
from the mass splitting gives more attraction in the $\Lambda\Lambda$ 
channel than in the $\Lambda N$ channel. Thus even though the 
$^1$S$_0$ interaction in the $\Lambda\Lambda$ channel may be weak 
(small scattering length), the additional attraction resulting from 
the coupling due to the baryon mass splitting could result in a scattering 
length that is comparable to the $nn$ scattering length as first 
suggested by Dover.\cite{D94} 

Let us now turn to the $\Lambda\Lambda$ interaction having 
established that the coupling between the $\Lambda\Lambda$ and $\Xi N$ 
channels is more important than is the case in the $S=-1$ channel. 
Since the experimental data put a constraint on the scattering amplitude 
that included the additional attraction due to the coupled channel 
nature of the problem, the $\Lambda\Lambda$ potential in free space, 
in the absence of coupling, is considerably stronger than is the case 
when the coupling is included.  However in the medium, the coupling is 
suppressed as a result of the fact that $\Lambda + \Lambda\rightarrow 
\Xi + N$ is Pauli blocked. From this we may conclude that while in 
free space 
\[
    \bra\,V_{\Lambda\Lambda}^{\rm eff}\,\ket_{\mbox{with coupling}}\ =\  
\bra\,V_{\Lambda\Lambda}^{\rm eff}\,\ket_{\mbox{no coupling}}\ ,
\]
in the medium
\[
\bra\,V_{\Lambda\Lambda}^{\rm eff}\,\ket_{\mbox{with coupling}}\ <\  
\bra\,V_{\Lambda\Lambda}^{\rm eff}\,\ket_{\mbox{no coupling}}\ . 
\]
In other words, if the $\Lambda\Lambda$ amplitude in free space is 
attractive enough to almost support a bound state, in the medium the 
Pauli principle suppresses this amplitude so that the effective matrix 
element of the $\Lambda\Lambda$ interaction is smaller than is the 
case for the $nn$ interaction.

The above qualitative analysis is based on the assumption that 
flavor $SU(3)$ gives a good estimate of the relative magnitude of 
matrix elements and the breaking of $SU(3)$ is only in the masses of 
the baryons. To see if these arguments are valid, in the 
next section we will report on the construction of a potential in the 
$S=-2$ channel based on the $SU(3)$ rotation of the Nijmegen model $D$ 
potential,\cite{N77} and then proceed in Sec.~\ref{sec.4} to 
use  the resulting potential in the calculation of the binding 
energy of $^{\ \ 6}_{\Lambda\Lambda}$He. Finally, in Sec.~\ref{sec.5} 
we will present some concluding remarks.

\section{The $\mathbf{S=-2}$ $\mathbf{BB}$ Potentials}\label{sec.3}

To perform an $SU(3)$ rotation on an OBE potential defined in $S=0,-1$ 
channels, we need to write the Lagrangian in terms of the baryon octet 
with the mesons as either a singlet or a member of an octet.
If the interaction is taken to be of the Yukawa type, the Lagrangian 
takes the form
\begin{equation}
{\cal L}_{int} = -\left\{ g_{\{1\}}[B^{\dag} B]_{\{1\}} M_{\{1\}} +  
            g_{\{8_s\}}[B^{\dag} B]_{\{8_s\}} M_{\{8\}} +
      g_{\{8_a\}}[B^{\dag} B]_{\{8_a\}} M_{\{8\}} \right\}\ , \label{eq:7}
\end{equation}
where $B$ and $M$ are the field operators for the baryon and mesons. 
In writing the above Lagrangian, which is a scalar, we have coupled the 
initial and final baryons to a flavor singlet or an octet. Since there
are two irreducible octet representations, we need a different coupling 
constant for each of the representations. This Lagrangian has one
coupling constant for each singlet meson $g_{\{1\}}$, and two coupling 
constants $g_{\{8_{s}\}}$ and $g_{\{8_{a}\}}$ for each meson 
octet. These coupling constants can then be determined by fitting the 
experimental data.

The Nijmegen potential\cite{N77} model $D$ takes for the exchange 
mesons the psedoscalar octet $\{\pi,\eta,\eta',K\}$, the vector octet 
$\{\rho,\phi,\omega,K^{*}\}$ and a scalar meson $\{\epsilon\}$. The 
masses of the mesons and baryons are taken from experiment, while the 
coupling constants are adjusted to fit the data in the $S=0,-1$. This, in 
principle, determines the long range part of the potential which should 
be described in terms of meson-baryon degrees of freedom. These same
coupling constants can be used to construct the OBE potential for
$S\le -2$. For the short range part of the interaction, model $D$ 
assumes a hard core, which is an additional parameter that could be 
different in the $S=0$ and $S=-1$ channels. The breaking of flavor
$SU(3)$ is due to the use of physical masses for the baryons and
mesons, and the choice of different hard core radii in the $S=0,-1$.

We expect the short range part of the $BB$ interaction to be 
expressed in terms of quark-gluon degrees of freedom. In a 
valence quark model description of this short range interaction the 
number of strange quarks increases as we go from $S=0$ to $S=-1$ and
then $S\le -2$. As the number of strange quarks increases, we expect
the role of the Pauli principle at the quark level to change. As a
result the short range part of the $BB$ interaction should be $S$ 
dependent, and cannot be determined from the $S=0,1$ channels at 
the meson-baryon level in terms of a flavor $SU(3)$ rotation of the
OBE potential. Thus to determine the $BB$ interaction in the $S=-2$ 
channel, we needs to introduce at least one additional parameter 
that determines the short range interaction. 

\begin{figure}[t]
\begin{minipage}[b]{.49\linewidth}
     \centerline{\includegraphics[scale=0.55]{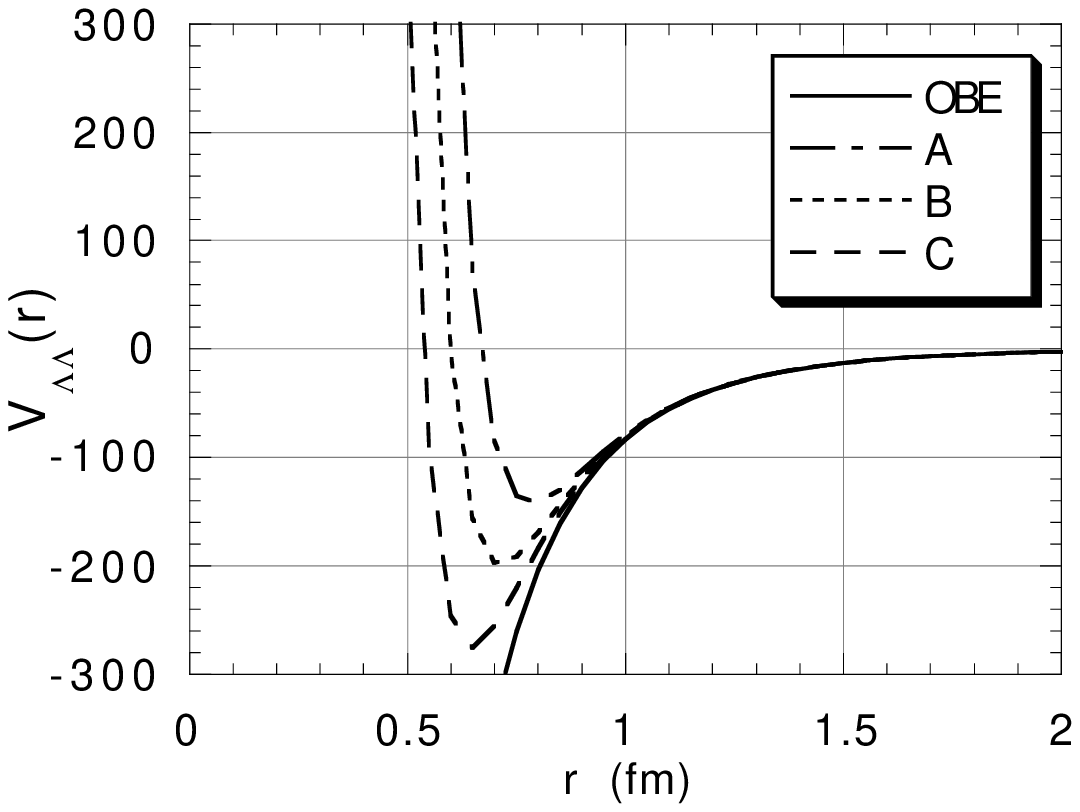}}
\end{minipage}\hfill
\begin{minipage}[b]{.49\linewidth}
     \centerline{\includegraphics[scale=0.55]{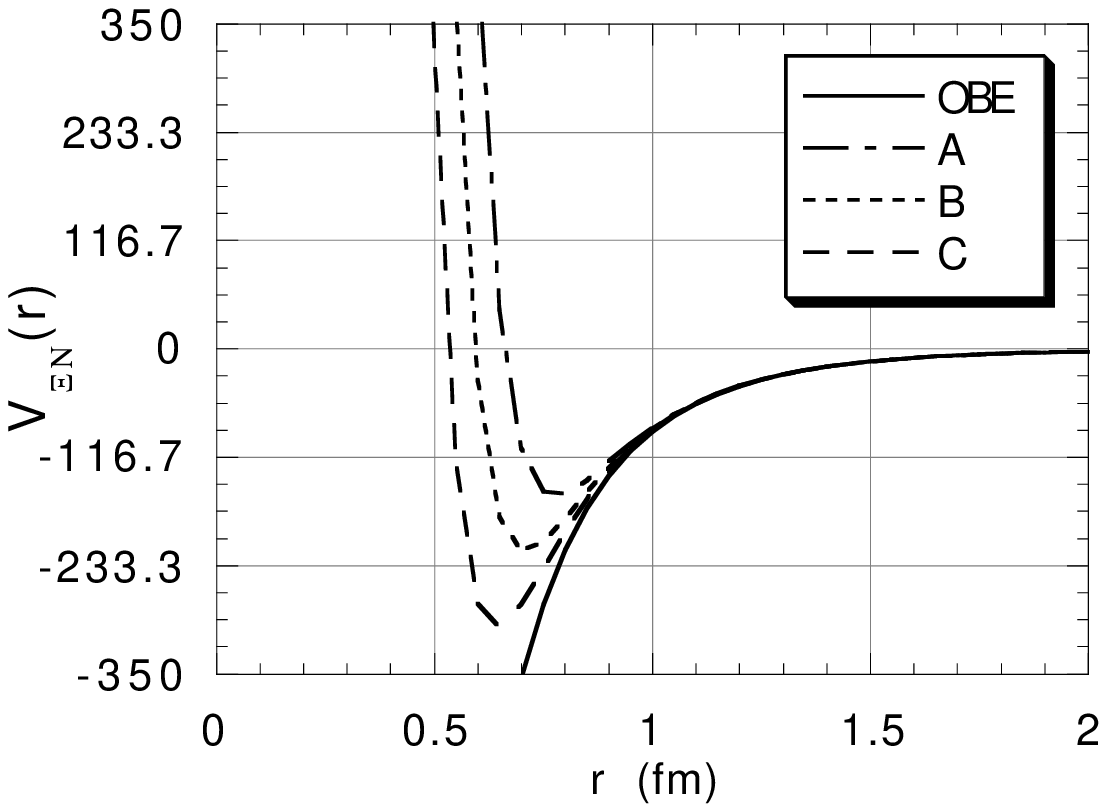}}
\end{minipage}
\caption{The $S=-2$, $^{1}$S$_{0}$ one-boson-exchange potential 
$V_{\Lambda\Lambda}$ and $V_{\Xi N}$.}\label{fig1}
\end{figure}

This procedure is followed by Carr {\it et~al}.\cite{C97} They only 
considered $S$-wave interaction and ignored the tensor interaction. Their 
potential for the exchange of the $i^{\rm th}$ meson is of the form
\begin{equation}
V_i(r) = V^{(i)}_c(r) 
       + \vec{\sigma}_1\cdot\vec{\sigma}_2\ 
       V^{(i)}_\sigma(r)                 \ ,        \label{eq:8}
\end{equation}
where the radial potential $V^{(i)}_\alpha, \alpha = c, \sigma$ for a 
mason of mass $m_{i}$ is taken to be
\begin{equation}
V_{\alpha}^{(i)}(r) = V_0^{(i)} \left[\frac{e^{-m_ir}}{m_ir} 
          - C\,\left(\frac{M}{m_i}\right)\, 
                     \frac{e^{-M r}}{M r}\right]
            \ ;\ \ \alpha = c,\sigma\ .            \label{eq:9}
\end{equation}
To guarantee a one parameter short range repulsion, a mass $M=2500$~MeV 
is chosen in all partial waves. Then the parameter $C$ determined the 
strength of the short range interaction. This new parameter $C$ is
constrained  to ensure that the potential for $r\ge 0.7$~fm is unchanged 
and the short range interaction is always repulsive. In Fig.~\ref{fig1}  
we present the $\Lambda\Lambda$ potential $V_{\Lambda\Lambda}$ and 
the $\Xi N$ potential $V_{\Xi N}$. In addition to the OBE potential
(solid line), the figure includes three potentials in which the
parameter $C$ is adjusted to support a bound state~(C), generate an 
an anti-bound state~(B), or have no bound state at all~(A).

To examine the role of the coupling between the $\Lambda\Lambda$ and 
$\Xi N$ channels in $^{\ \ 6}_{\Lambda\Lambda}$He, we need to 
include this coupling exactly. This can be achieved most simply within a 
model in which $^{\ \ 6}_{\Lambda\Lambda}$He is taken to 
be a $\Lambda\Lambda\alpha$ three-body system described by the 
Faddeev equations. These equations can be reduced to coupled one 
dimensional integral equations if the interactions are assumed to be 
separable. For this Carr~{\it et~al}\,\cite{C97} generated a set of separable 
potentials that give the same effective range parameters as meson
exchange potentials A, B and C. These are referred to as SA, SB and
SC. The effective range parameters for both the local and the separable 
potentials are given in Table~\ref{table1}. Here we observe that the 
potentials B and SB have a scattering length that is comparable to the 
$nn$ scattering length. 

\begin{table}[b]
    \begin{center}
	\caption{The effective range parameters and binding energy of the 
	two-body $\Lambda\Lambda-\Xi N$ system for the local
        potentials A, B and C and the equivalent separable potentials 
        SA, SB and SC.}\label{table1}
\begin{tabular}{|c|cc|cc|c|} \hline\
 Pot. & $a_{\Lambda\Lambda}$ & $r_{\Lambda\Lambda}$ & $a_{\Xi N}$ & $r_{\Xi
N}$ & B.E. (MeV) \\ \hline
 & & & & & \\
    A    & -1.91 & 3.36 & -2.12-0.75i & 3.45-0.45i & UB   \\
   SA    & -1.90 & 3.33 & -2.08-0.81i & 3.44-0.22i & UB   \\
 & & & & & \\
    B    & -21.1 & 1.86 & -2.05-6.53i & 2.12-0.21i & UB   \\
   SB    & -21.0 & 2.54 & -2.07-6.52i & 2.62-0.15i & UB   \\
 & & & & & \\
    C    & 7.82  & 1.41 & 3.08-5.26i  & 1.74-0.144i& 0.71 \\
   SC    & 7.84  & 1.48 & 3.05-5.28i  & 1.45+0.074i& 0.71 \\
\hline
\end{tabular}
\end{center}
\end{table}

\section{Binding Energy $\mathbf{^{\ \ 6}_{\Lambda\Lambda}}$He}\label{sec.4}

In  reducing the six body problem of $^{\ \ 6}_{\Lambda\Lambda}$He to 
a three-body problem we gain the advantage of being able to solve 
the problem exactly using the Faddeev equations. In particular, the 
treatment of the coupled channel $\Lambda\Lambda-\Xi N$ is included 
in full. The main disadvantage of this three-body model is the 
limitation in the handling of the Pauli principle. This is most 
pronounced when the two $\Lambda$s convert to $\Xi N$, as the resulting 
nucleon should be in an antisymmetric state with respect to the 
nucleons in the $\alpha$ particle. Since the latter is treated as an 
elementary particle with no internal degrees of freedom, it is not 
possible to guarantee that the Pauli principle is satisfied. This effect is 
most pronounced when the nucleon and $\alpha$ particle are in 
relative $S$-wave. There are two ways of overcoming this problem in 
an approximate way. The first is to assume that the $S$-wave 
interaction supports a bound state in the absence of the Pauli
exclusion principle. We then can remove this bound state from the 
spectrum of the Hamiltonian without any modification to the phase 
shifts. The resultant amplitude has the Pauli forbidden state
removed. The second approach is to take the $S$-wave $N-\alpha$ 
interaction to be repulsive. The two methods give almost identical 
results for the binding energy of $^{6}$Li.\cite{L78} 
Carr {\it et al}\,\cite{C97} used the latter approximation 
as it results in a rank one potential and therefore a simpler set of 
Faddeev equations for the three-body model of 
$^{\ \ 6}_{\Lambda\Lambda}$He 

The second complication in turning $^{\ \ 6}_{\Lambda\Lambda}$He into a 
three-body problem is the need to know the $\Lambda-\alpha$ and 
$\Xi-\alpha$ interactions. These interactions should be consistent with 
the two-body $BB$ interaction. Here again, Carr {\it et al}\,\cite{C97} 
resorted to simple models for these interactions. For the case of the 
$\Lambda-\alpha$ interaction, they made use of the $G$-matrix 
parametrization\cite{Y85} of the Nijmegen model $D$ potential to 
fold this interaction with $^{4}$He density, while for the 
$\Xi-\alpha$ interaction they employed a Wood-Saxon optical 
potential.

Finally, since the conversion of the $\Lambda\Lambda$ to $\Xi N$ 
releases $\approx 25$~MeV of energy, that energy could go into the 
excitation of the $\alpha$ particle. Although it is possible to include 
such excitation by assuming the $\alpha$ particle can be in one of two 
states with mass separation comparable to the energy of the first 
excited state of $^{4}$He, Carr {\it et~al}\,\cite{C97} chose not to 
include this excitation as the effective coupling to the $\Xi 
N\alpha$ channel is suppressed because the $S$-wave $\alpha N$ 
interaction has been taken to be repulsive, i.e. because of Pauli
blocking.  

To examine the relative importance of the coupling between the 
$\Lambda\Lambda$ and $\Xi N$ in free space, and in the nuclear 
medium where the Pauli blocking plays an important role, we 
consider three models for the binding energy of 
$^{\ \ 6}_{\Lambda\Lambda}$He. These are:
\begin{enumerate}
  \item In model 1 we include the full coupling between the 
  $\Lambda\Lambda$ and the $\Xi N$. This reduces the three-body model 
  for $^{\ \ 6}_{\Lambda\Lambda}$He to the three-body coupled channel 
  problem $\Lambda\Lambda\alpha-\Xi N\alpha$, which can be solved 
  exactly to give the binding energy for the three-body system under 
  consideration. This is an exact solution to the problem within the
  limitations stated above. Here we note that this system is not a
  pure $\Lambda\Lambda\alpha$ system, but has a certain probability 
  of being in a $\Xi N\alpha$ state.
  \item In model 2 we include the coupling between the $\Lambda\Lambda$ 
  and $\Xi N$ channels in calculating the two-body $T$-matrix that 
  is the input to the Faddeev equations, but carry out the 
  three-body calculation within the $\Lambda\Lambda\alpha$ 
  space only. This is equivalent to setting $T_{\Lambda\Xi}=0$,
  which gives us a measure of the importance of the coupling between 
  the $\Lambda\Lambda$ and $\Xi N$ in nuclear matter, and the effect 
  of the Pauli blocking. In this case the probability of finding the
  system in $\Xi N\alpha$ state is zero.
  \item Finally, in model 3 we assume that the coupling between the 
  $\Lambda\Lambda$ and $\Xi N$ can be neglected at the two-body level 
  in free space, i.e. $V_{\Lambda\Xi}=0$. In this case we construct 
  a two-body $\Lambda\Lambda$ interaction that gives the same
  $\Lambda\Lambda$ effective range parameters as the coupled channel 
  problem. Thus for model B, the $\Lambda\Lambda$ potential, in the
  absence of coupling, has an anti-bound state and a scattering length 
  comparable to the $nn$ scattering length.
\end{enumerate}

\begin{table}[b]
\begin{center}
    \caption{The binding energy of $^{\ \ 6}_{\Lambda\Lambda}$He for 
    the three models under consideration and for the three 
    interactions SA, SB and SC.}\label{table2}
\begin{tabular}{|c|c|c|c|c|} \hline 
 & & & & \\
 & $a_{\Lambda\Lambda}$ & BE(model 1) & BE(model 2) &
BE(model 3) \\
 & &( $\Lambda\Lambda\alpha-\Xi N\alpha$ ) & 
 ( $\Lambda\Lambda\alpha$; $T_{\Lambda\Xi}=0$ ) &
 ( $\Lambda\Lambda\alpha$; $V_{\Lambda\Xi}=0$ ) \\ 
 & (fm)     & (MeV)    & (MeV)    &  (MeV) \\  \hline
 & & & &  \\
 SA & -1.91  &  9.738   & 9.5078   & 10.007 \\ 
 & & & &  \\
 SB & -21.1  & 12.268   & 11.606   & 14.134 \\ 
 & & & &  \\
 SC & 7.82  & 15.912   & 14.533   & 17.842 \\ 
 & & & &  \\
 Exp. &  & 10.9 $\pm$ 0.6   &      &       \\ 
 & & & & \\ \hline
\end{tabular}
\end{center}
\end{table}

In Table~\ref{table2} we present the results of 
Carr {\it et~al}\,\cite{C97} for the above three models using the 
potentials SA, SB and SC for the $\Lambda\Lambda-\Xi N$ interaction. 
We observe that: 
\begin{itemize}
\item The experimental result of $10.9 \pm 0.6$ lies between the 
``exact'' results (model 1) for potentials SA and SB, and closer 
to SB. This suggests that within the present model the short range 
parameter $C$ should be such that the $\Lambda\Lambda$ scattering 
length is large and comparable to the $nn$ scattering length if we 
are to reproduce the binding energy of $^{\ \ 6}_{\Lambda\Lambda}$He. 
Therefore we can use the binding energy of one of the 
$\Lambda\Lambda$ hypernuclei to fix the short range part of the 
interaction in the $S=-2$ channel and then proceed to use the resulting 
potential for other $S=-2$ hypernuclei.
\item If we now compare the results of model 1 and model 2 in 
Table~\ref{table2}, 
we observe that the contribution of $T_{\Lambda\Xi}$ is small 
($\approx 0.5$~MeV) if our potential would give a binding energy 
comparable to the experimental result. This relatively small 
reduction in binding energy when the $\Lambda\Lambda-\Xi N$ 
coupling is included is a measure of the importance of Pauli 
blocking, even in a nucleus as light as $^{\ \ 6}_{\Lambda\Lambda}$He. 
This small change in binding energy due to coupling in the nuclear 
medium does in part justify the exclusion of the excitation of 
the $\alpha$ particle in the present models.
\item On the other hand, if we compare the results of model~1 and 
model~3 in Table~\ref{table2}, we may conclude that turning off the  
coupling between the $\Lambda\Lambda$ and $\Xi N$ channels 
at the two-body level, in free space, is in fact a very poor 
approximation. The difference of $\approx 2$~MeV between model~1 
and model~3 is a substantial fraction of the value of the 
$\Lambda\Lambda$ matrix element (a change of $\approx 30\%$). 
This difference is, to lowest order, the contribution of the 
coupling between the $\Lambda\Lambda$ and $\Xi N$ to the attraction 
in the  effective $\Lambda\Lambda$ matrix element. To see this, 
we write the binding energy of $^{\ \ 6}_{\Lambda\Lambda}$He as 
the matrix element of the Hamiltonian, i.e.
\begin{equation}
   -  B.E. = \bra H \ket = K.E. + 2\bra V_{\Lambda\alpha}\ket 
         + \bra V^{\rm eff}_{\Lambda\Lambda}\ket  \ ,  \label{eq:10}
\end{equation}
where $\bra V^{\rm eff}_{\Lambda\Lambda}\ket$ is the effective 
$\Lambda\Lambda$ matrix elements in the medium. If there is no 
coupling between the $\Lambda\Lambda$ and $\Xi N$ channels in free 
space then
\begin{equation}
\bra V^{\rm eff}_{\Lambda\Lambda}\ket = 
   \bra V_{\Lambda\Lambda}\ket_{\mbox{no coupling}} \quad
   \mbox{for}\quad V_{\Lambda\Xi}=0      \ .     \label{eq:11}
\end{equation}
On the other hand, the inclusion of the coupling between the 
$\Lambda\Lambda$ and the $\Xi N$ in free space reduces the strength 
of the $\Lambda\Lambda$ potential, $V_{\Lambda\Lambda}$, so 
that the $\Lambda\Lambda$ potential and the additional attraction 
due to the coupling results in an effective matrix element that is 
fixed by experiment. If we assume that the Pauli blocking removes the 
contribution of the coupling between the $\Lambda\Lambda$ and $\Xi N$ 
in the nuclear medium, i.e. the difference between the result of 
model~1 and model~2 in Table~\ref{table2} is zero, then
\begin{equation}
 \bra V^{\rm eff}_{\Lambda\Lambda}\ket = 
   \bra V_{\Lambda\Lambda}\ket_{\mbox{with coupling}} 
 \quad\mbox{for}\quad V_{\Lambda\Xi}\neq 0\ . \label{eq:12}
\end{equation}
But from Eq.~(\ref{eq:10}) we know that the difference between the 
binding energy with and without coupling between the $\Lambda\Lambda$ 
and the $\Xi N$ channels in free space is the contribution of the 
coupling to the attraction, i.e.
\begin{eqnarray}
    BE(\mbox{model 1}) - BE(\mbox{model 3}) &\approx& 
    - \bra V_{\Lambda\Lambda}\ket_{\mbox{with coupling}}
    + \bra V_{\Lambda\Lambda}\ket_{\mbox{no coupling}} \nonumber \\
    & & \nonumber \\
    &\approx& \frac{|\bra\Lambda\Lambda |V|N\Xi\ket|^2}
                        {\Delta E_{\Lambda\Lambda}}
    \approx -2\ \mbox{MeV}\ . \label{eq:13}
\end{eqnarray}
This result is approximately the difference between the $nn$ and 
$\Lambda\Lambda$ matrix elements in light nuclei.
\end{itemize}
This analysis basically establishes the fact that it is essential to 
include the coupling between the $\Lambda\Lambda$ and $\Xi N$ 
channels in free space even though the coupling is suppressed in the 
medium due to Pauli blocking.

\section{Conclusion}\label{sec.5}

From the above analysis based on the validity of flavor $SU(3)$ 
for matrix elements of the potential, we  can understand how the 
coupling resulting from mass differences between the members of 
the baryon octet could justify the fact that 
$- \bra V_{\Lambda\Lambda}\ket > - \bra V_{\Lambda N}\ket$. 
In fact flavor $SU(3)$ suggests that the coupling in the $S=-2$ 
is substantially more important than is the case in the $S=-1$ 
channel. This is a result of the fact that: (i)~The 
$\Lambda\Lambda$ threshold and the $\Xi N$ threshold are closer than 
the $\Lambda N$ and $\Sigma N$ thresholds. (ii)~The coupling matrix 
element for the the transition from $\Lambda\Lambda$ to $\Xi N$ is 
larger than the coupling matrix elements for the transition from 
$\Lambda N$ to $\Sigma N$.

With the help of flavor $SU(3)$ rotation we showed how we can 
construct the long range part of the potential based on OBE for 
$S\le -2$ given that the meson baryon coupling constants are fixed by 
the data in the $S=0,-1$ channels. However, this flavor $SU(3)$ 
rotation does not determine the short range part of the interaction. 
In fact valence quark models for the $BB$ system would give an $S$ 
dependence to the short range data, and in an OBE model this requires 
additional data, e.g. the binding energy of $\Lambda\Lambda$ 
hypernuclei.

Using this $SU(3)$ rotation Carr~{\it et al}\,\cite{C97} constructed an 
extension of the Nijmegen model D potential to the $S=-2$ channel with 
a soft cut-off that does not change the potential outside of 0.7~fm. 
Finally, Carr~{\it et~al}\,\cite{C97} were able to show that a 
$\Lambda\Lambda$ scattering length comparable to the $nn$ scattering 
length gives approximately the correct binding energy 
for $^{\ \ 6}_{\Lambda\Lambda}$He. This suggests that the 
experimental binding energy of $^{\ \ 6}_{\Lambda\Lambda}$He is 
consistent with the strength of the $BB$ potential as predicted by 
flavor $SU(3)$ rotation of the OBE potentials. More important it 
confirms the prediction of Dover\cite{D94} that the $\Lambda\Lambda$ 
scattering length is comparable to the $nn$ scattering length in the 
$^{1}$S$_{0}$.

With the inclusion of coupling between the $\Lambda\Lambda$ and $\Xi 
N$ channel there is no need to include repulsive three-body forces 
to calculate the correct binding energy for 
$^{\ \ 6}_{\Lambda\Lambda}$He. This is 
a result of the fact that the Pauli blocking of the process 
$\Lambda\Lambda\rightarrow\Xi N$ in the nuclear medium reduces the 
effective attraction in the $\Lambda\Lambda$ channel as compared to 
the attraction in free space. As a result the $\Lambda\Lambda$ matrix 
element in light nuclei is less than the $nn$ matrix element by about 
2~MeV, which is consistent with the experimental results from 
$\Lambda\Lambda$ hypernuclei. This is despite the fact that the 
scattering length for the $\Lambda\Lambda$ is comparable to the $nn$ 
scattering length.

Finally, we can understand the result that 
\begin{equation}
    -\bra V_{nn}\ket > -\bra V_{\Lambda\Lambda}\ket >
    -\bra V_{\Lambda N}\ket                 \ ,  \label{eq:14}
\end{equation}
on the basis of the coupled channel nature of the $BB$ interaction in 
the $S=-1,-2$ channels resulting from flavor $SU(3)$ breaking in the 
mass of the baryon octet.

\section*{Acknowledgments}

The author would like to thank the Australian Research Council for 
their financial support. He is indebted to his collaborators S.B. Carr 
and B.F. Gibson for many enlightening discussions and helpful 
suggestions.

\end{document}